\documentstyle[twoside,fleqn,espcrc2]{article}

\newcommand{\beq}{\begin{equation}}
\newcommand{\eeq}{\end{equation}}
\newcommand{\beqa}{\begin{eqnarray}}
\newcommand{\eeqa}{\end{eqnarray}}
\newcommand{\phil}{\phi_{\rm L}}
\newcommand{\eatspace}{\vspace{-0.1in}}
\newcommand{\eatless}{\vspace{-0.06in}}

\title{Atomic Bose Condensation and the Lattice}

\author{Guy D. Moore\address{Department of Physics, University of
        Washington, Seattle, WA 98195, USA}}
       
\begin{document}
\input epsf.tex

\begin{abstract}

I show how interaction corrections to the Bose condensation temperature
of an atomic gas can be computed using a combination of perturbative
effective field theory and lattice techniques.

\end{abstract}

\maketitle

\section{Introduction}
\eatless

This is a lattice talk, but at first it won't seem like it; it will be
about atomic systems, and effective field theories.  But studying the
effective description of the atomic system will lead to a question
suited to lattice techniques.
I am summarizing 3 papers \cite{paper1,paper2,paper3}, all with
Peter Arnold; the middle one contains all the lattice stuff.

Many systems exhibit Bose condensation; for instance, liquid Helium,
Cooper pairs in superconductors, and arguably nucleon Cooper pairs 
in large nuclei.
These examples condense because of interactions; none exhibit Bose 
condensation as originally proposed by Bose and Einstein, where
condensation occurs exclusively because 
the temperature is so low that
the only place to store the bosons is in the ground state.

The interest in atomic Bose condensation mainly arises because this is
what {\em does} happen here; mutual interactions are not that
important, the dominant reason for condensation is that there is not
enough heat to store the density of atoms present in the excited
states.  In fact, interactions {\em must} be weak in an atomic Bose
condensate; otherwise it would condense into a solid.

The beauty of small interactions is it means we can calculate;
the coupling is weak enough that perturbatively based
``field theorist'' techniques, like effective field theories and
loopwise expansions, can be used.

The question I will address in this talk is, how do interactions modify
the relation between density and the temperature of the Bose condensation
transition?  For simplicity, and to follow the literature,
I consider a homogeneous system.  (This also makes the question well
posed.)  It means that the problem addressed is
a little artificial and has less contact with experiment than we would
like; but nevertheless since there is already a surprisingly big
literature on the problem as posed, it is worth addressing.

\eatless
\section{Effective theories}
\eatless

The right effective description of a trapped atomic gas of bosonic
atoms, all in the same spin state, is a second quantized Schr\"{o}dinger
equation:
\eatspace
\begin{eqnarray}
{\cal L} & = & \int d^3 x \;
	\Psi^\dagger \!\left( i \hbar \, \partial_t + \frac{\hbar^2}{2m} 
	\nabla^2 - V(x) \right) \! \Psi \nonumber \eatless \\ & & 
	- \int d^3 x \, d^3 y 
	\; V(x,y) \: \Psi^\dagger \Psi(x) \, \Psi^\dagger \Psi(y) \, .
\eatspace
\end{eqnarray}
The atomic number density is $\Psi^\dagger \Psi$, with a
subtraction implied so the density vanishes in vacuum.  
Call this Theory I.  Technically, Theory I is already an
effective theory, containing no information about atomic structure; it
is only valid on length scales larger than the atomic size.
Since we are concerned with a dilute system we will be interested in
scales large compared with the scattering length, in which case the nonlocal
interaction term should be replaced with a local term plus higher
dimension corrections.  Furthermore we are interested in the behavior at
finite temperature, and will only ask about static properties (namely
the number density), so we can go to the Matsubara formalism, with
Euclidean periodic time.  I work in the grand canonical ensemble, so I
introduce a chemical potential $\mu$, and because the system is
homogeneous, $V(x)$ can be dropped.  I will also set $\hbar = 1$ from
now on.  The appropriate effective theory is therefore
\eatspace
\begin{eqnarray}
S_{\rm II} & \!\!\! = \!\!\! & \int_0^\beta d\tau \int d^3 x 
	{\cal L}_{\rm E} \, , \\
{\cal L}_{\rm E} & \!\!\!\! = \!\!\!\! & \Psi^\dagger 
	\!\left( \! - \partial_\tau
	- \frac{\nabla^2}{2m} - \mu \! \right) \! \Psi 
	+ \frac{2\pi a_{\rm sc}}{m} (\Psi^\dagger \Psi)^2 
	\, ,
\eatspace
\end{eqnarray}
plus high dimension operators.
The number density is still $\Psi^\dagger \Psi$, with its zero
temperature value subtracted.  This effective theory requires
regularization, and I use $\overline{\rm MS}$ dimensional
regularization.  $a_{\rm sc}$ is the scattering length; we could derive
it by matching to Theory I, but we rather define it to be the 
coefficient shown.  There are also high dimension operators, containing
either more fields or derivatives; but at the 
order of interest they can all be dropped.  I will call this Theory II.

Theory II contains three interesting length scales $\lambda$:
\begin{enumerate}
\item{$\lambda \sim a_{\rm sc}\sim 5{\rm nm}$, where it breaks down;}
\item{$\lambda \sim \lambda_{\rm T} \equiv \sqrt{2\pi/mT}\sim 
      100$--$300{\rm nm}$, 
      the thermal scale, where nonzero Matsubara frequencies decouple and
      the behavior becomes effectively 3D;}
\item{$\lambda \sim \lambda_{\rm T}^2 / a_{\rm sc}$, the scale
      were the behavior becomes strongly coupled.}
\end{enumerate}
The physics of Bose condensation is associated with the third scale,
where the nonzero Matsubara modes have decoupled and the physics is
effectively 3D.  Therefore we can write a still simpler effective theory
which contains the physics of Bose condensation, which I call Theory
III:
\eatspace
\begin{eqnarray}
S_{\rm III} & \!\!\! = \!\!\! & \int d^3 x \; {\cal L}_{\rm 3D} \, \\
{\cal L}_{\rm 3D} & \!\!\! = \!\!\! & \frac{1}{2} \phi 
	\! \left( -\nabla^2 + r \right) \! 
	\phi + \frac{u}{24} (\phi^2)^2 \, ,
\eatspace
\end{eqnarray}
with $\phi$ a two component real field (the two components correspond to
the real and imaginary parts of $\Psi$).
Again there are in principle high dimension terms, but we won't
need them.  The possibility of such matching was realized by Baym
et.~al.~\cite{Baym}. 

The relation between $\mu$, $a_{\rm sc}$, and $\langle \Psi^\dagger
\Psi \rangle$ in Theory II and $r$, $u$, and $\langle \phi^2 \rangle$ in
Theory III is determined by matching between the theories.
The coupling is weak at $\lambda = \lambda_{\rm T}$, so the matching may
be done perturbatively.
At tree level, $u=96\pi^2 a_{\rm sc}/\lambda_{\rm T}^2$,
$\langle \phi^2 \rangle = \langle \Psi^\dagger \Psi \rangle /m$, 
and $r = -2m\mu$.  The operator 
$\phi^2$ mixes with the identity. A one loop matching calculation of the
mixing gives the free theory value for the particle number,
\eatless
\begin{equation}
\label{eq:n_is}
n_{\rm free} = \zeta(3/2) \; \lambda_{\rm T}^{-3} \, .
\eatspace
\end{equation}
To find the $O(a_{\rm sc}/\lambda_{\rm T})$ corrections to the number
density, we need to find this mixing to two loops, $r$ to one loop, and
the other quantities at tree level.
We can do even better, and determine the $O(a_{\rm
sc}^2 / \lambda_{\rm T}^2)$ corrections, by going to one higher loopwise
order in every variable.  This is done in \cite{paper3}.
We cannot go further without including high dimension operators, so we
will stop at this order.

Besides the matching calculation,
we also need the critical values of two quantities 
in Theory III.  At linear order in 
$a_{\rm sc}$ we need the value of $\langle \phi^2 \rangle$,
evaluated at the critical value $r_{\rm crit}$ (the critical $\mu$).
At second order in
$a_{\rm sc}$, we also need to know the critical value $r_{\rm crit}$.
$r_{\rm crit}$ is also needed to find the total number of atoms in a
wide but finite trap \cite{ArnoldTomasik}.

\section{The lattice}

The values of $\langle \phi^2 \rangle$ and $r_{\rm crit}$
are nonperturbative.  Theory III is the 3-D O(2) model,
and its infrared behavior falls in the x-y universality class.  The
infrared behavior is strongly coupled.  Furthermore, the quantities of
interest, $\langle \phi^2 \rangle$ and $r_{\rm crit}$, are 
not universal.  They cannot, to our knowledge, be reliably
determined using known analytic techniques (perturbation theory,
$\epsilon$ expansion, high temperature expansion, etc.).  We turn
instead to a lattice determination.

We will refer to the lattice regularization of Theory III as Theory IV.
$\phi$, a two component real field, is discretized on a 3D cubic lattice
of spacing $a_{\rm latt}$, with action
\begin{eqnarray}
S_{\rm IV} \!\!\! & = & \!\!\! a_{\rm latt}^3 \sum 
	{\cal L}_{\rm Latt} \, , \\
{\cal L}_{\rm Latt} \!\!\! & = & \!\!\! \frac{1}{2} \phil 
	\! \left( -Z_{\phi} \nabla_{\rm L}^2 + r_{\rm L} \right)
	\phil + \frac{u_{\rm L}}{24} (\phil^2)^2 \, .
\end{eqnarray}
Relating lattice parameters and Theory III parameters,
including $\langle \phil^2 \rangle$, requires another matching
calculation, which is identical in structure to the matching performed
between Theories II and III; it can be done by lattice perturbation
theory.  In \cite{paper2} we perform this matching to the same, and in
some cases higher, loop order, as we used in matching Theories II and
III.  The matching is an expansion in $a_{\rm latt} u$, which means we
want $a_{\rm latt}$ to be relatively small.  However we also want to be deep
in the critical scaling regime, requiring a large physical 
volume.  To improve the
matching to Theory III we use an improved lattice Laplacian, containing
anti-ferromagnetic next-nearest couplings (analogous to using the
Symanzik rather than Wilson action in pure glue QCD).  This frustrates
the cluster algorithm, but the multi-grid algorithm \cite{multigrid} is
available and proves quite efficient for Monte-Carlo use.

We determine the critical value $r_{\rm crit}$ by the method of Binder
\cite{Binder}, which accelerates the large volume convergence;
extrapolation of $r_{\rm crit}$ to infinite volume is well behaved.  Our
matching procedure leaves linear in $a_{\rm latt}$ errors in $r_{\rm
L}$, so this extrapolation is less trivial:

\begin{figure}[h]
\vspace{-0.9in}
\includegraphics{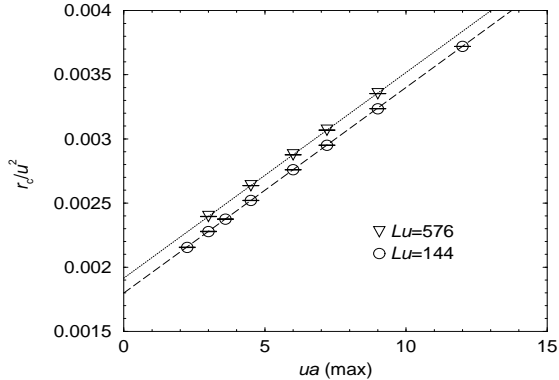}
\vspace{2.2in}
\caption{Zero spacing extrapolation for $r_{\rm crit}$.}
\end{figure}

\vspace{-0.3in}
The result is that, at renormalization point $\overline{\mu} = u/3$, 
$r_{\rm crit}/u^2 = 0.0019201(21)$ \cite{paper2}.

\begin{figure}[tbh]
\vspace{-0.8in}
\includegraphics{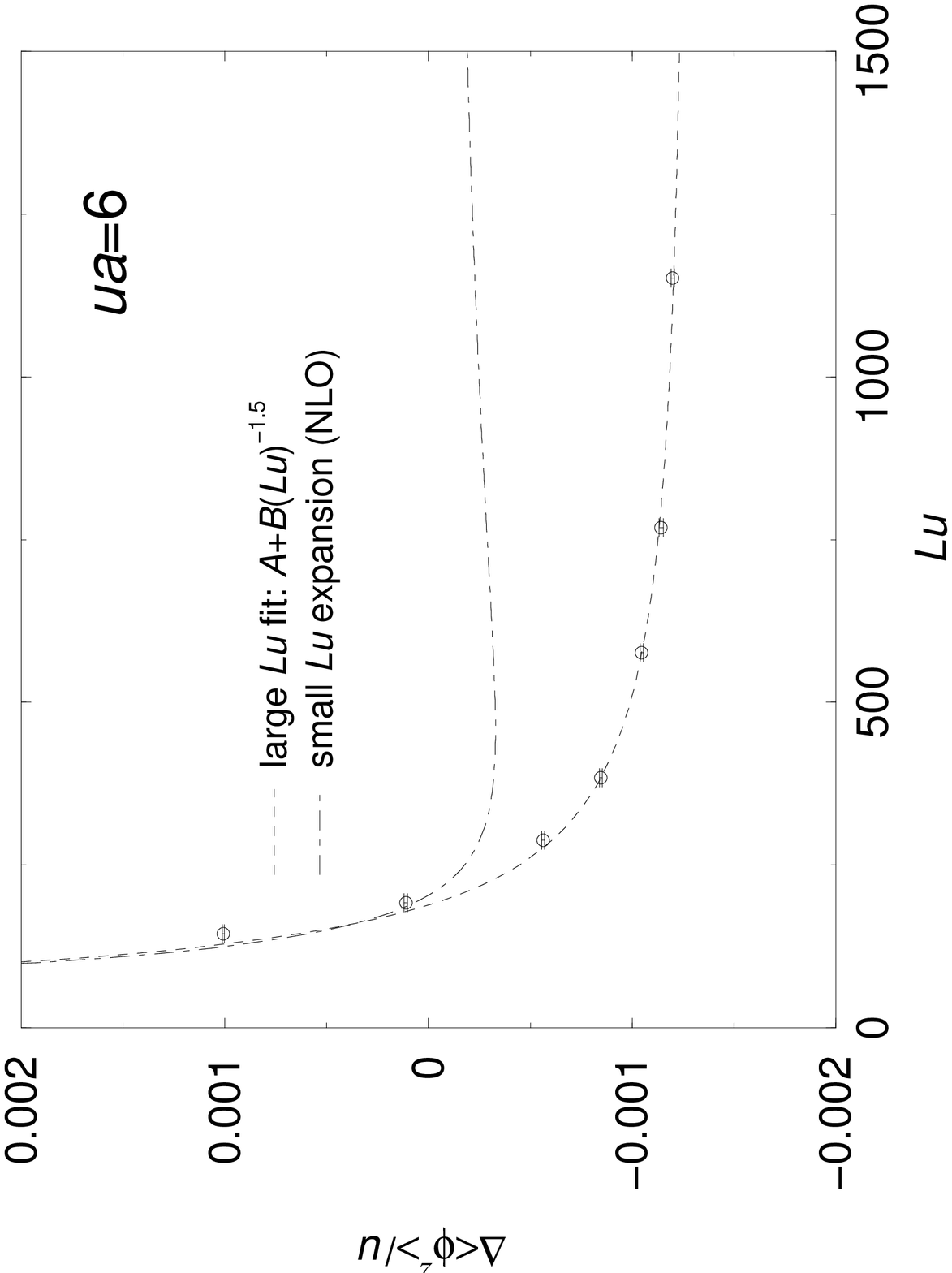}
\vspace{2.15in}
\includegraphics{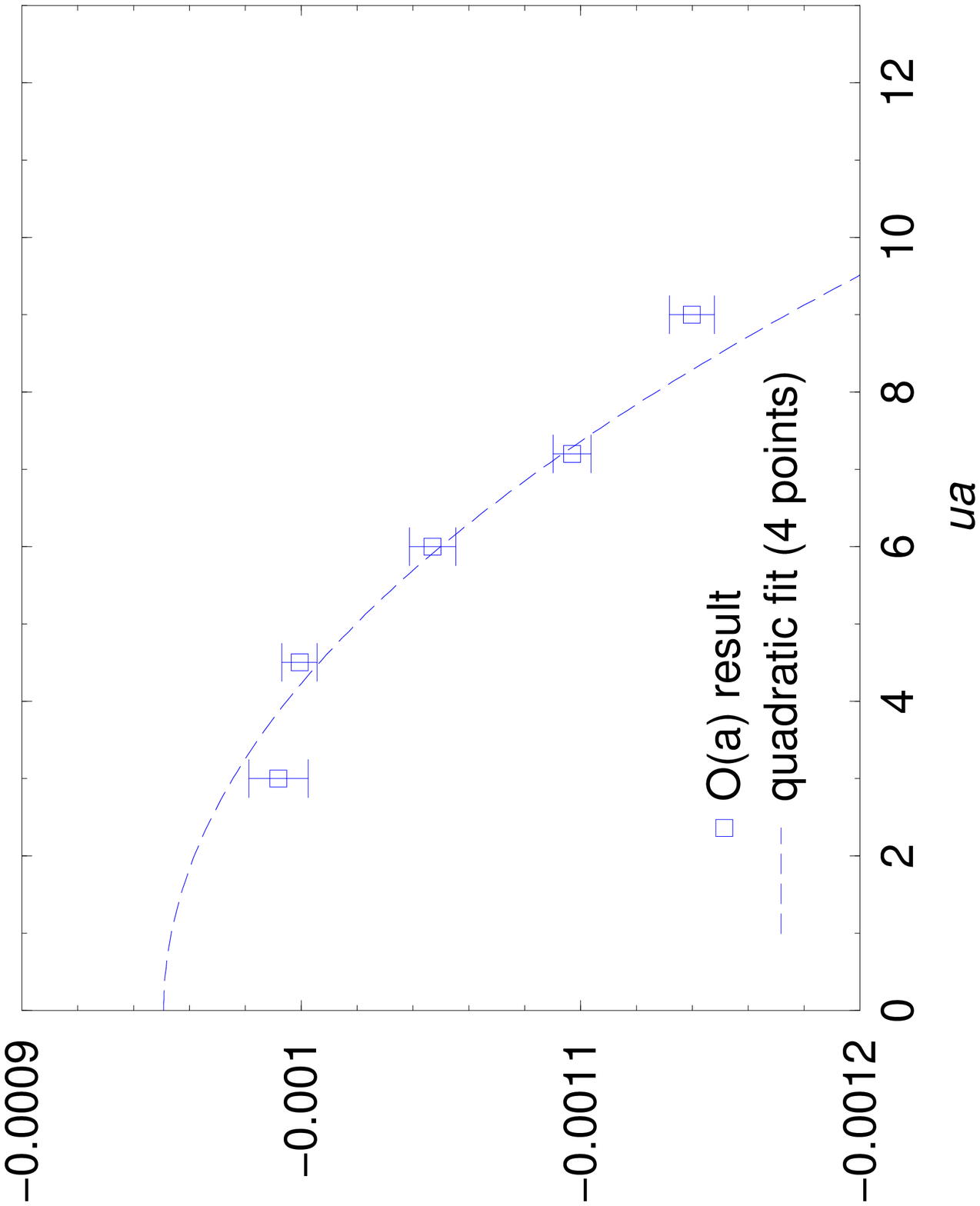}
\vspace{2.3in}
\caption{\label{fig2} Top:  volume behavior, fixed $a_{\rm latt}$.  
Bottom:  $a_{\rm latt}$ behavior, fixed volume.}
\end{figure}

For $\langle \phi^2 \rangle$ the large volume extrapolation is tougher,
and requires a careful accounting of the expected critical behavior.
However, the matching has eliminated more of the lattice spacing correction,
see Fig.~\ref{fig2}.  After a double extrapolation, we find
$\langle \phi^2_{\rm 3D} \rangle /u = -0.001198(17)$ \cite{paper2}.

Putting these results together with the results of the matching
calculation between Theory II and Theory III, we determine the
correction to $T_{\rm c}$, at fixed number density $n$, to second order
in the scattering length.  Defining $\rho=a_{\rm scatt} n^{1/3}$,
\begin{eqnarray}
T_{\rm c} \!\!\! & = & \!\!\! T_0 \left( 1 + c_1 \rho
	+ \left( c_2' \ln(\rho) + c_2'' \right) \rho^2  \right) \, , \\
c_1 \!\!\!&=&\!\!\! 1.32(2) \, , \quad
c_2' = 19.75 \, , \quad
c_2'' = 75.7(4) \, ,
\end{eqnarray}
where $T_0$ is determined by inverting Eq.~(\ref{eq:n_is}).

\vspace{-0.3in}

\eatspace

\end{document}